# A 3D APPARENT HORIZON FINDER[*]

JOSEPH LIBSON[1], JOAN MASSÓ[1,2], EDWARD SEIDEL[1], WAI-MO SUEN[3]
[1] *National Center for Supercomputing Applications*
*605 E. Springfield Ave., Champaign, Illinois 61820*
[2] *Departament de Física, Universitat de les Illes Balears,*
*E-07071 Palma de Mallorca, Spain*
[3] *McDonnell Center for the Space Sciences, Department of Physics*
*Washington University, St. Louis, Missouri, 63130*

ABSTRACT
We report on an efficient method for locating the apparent horizon in numerically constructed dynamical 3D black hole spacetimes. Instead of solving the zero expansion partial differential equation, our method uses a minimization procedure. Converting the PDE problem to minimization eliminates the difficulty of imposing suitable boundary conditions for the PDE. We demonstrate the effectiveness of this method in both 2D and 3D cases. The method is also highly parallelizable for implementation in massively parallel computers.

## 1. Introduction

The apparent horizon (AH) is a crucial characteristic of a black hole spacetime. For numerically constructed black hole spacetimes, while there are efficient methods to find the AH in a wide variety of 2 dimensional situations,[1] there is as yet no satisfactory method for the 3D cases.

The by now standard method of finding AH, at least for lower dimensional cases, is to solve the zero expansion equation:

$$\Theta = s^i{}_{|i} - K + K_{ij}s^i s^j = 0 \quad , \tag{1}$$

where $s^i$ is the outward spatial normal vector of the AH on the constant time slice at which the AH is to be found, and $K$ is the trace of the extrinsic curvature $K_{ij}$ of that slice. This equation can be easily integrated to determine the AH in spacetimes with symmetries, as the symmetries often provide boundary conditions at the edges of the computational domain (e.g., the axis and the equator). However, for a general 3D spacetime, as the AH is topologically a 2-sphere, one immediately faces the difficulty of imposing suitable boundary conditions even for the starting of the integration, let alone the devising of an efficient method for solving this nonlinear partial differential equation (PDE).

Our method for handling this problem of finding the AH surface in a general 3D spacetime is to convert the PDE problem to a minimization problem, the solution of which is well understood. This can be achieved by:

1. Express a surface $F(x_i) = 0$, which is topologically a 2-sphere living on a constant time slice, in terms of the symmetric trace free (STF) tensors,[2] or

---

other suitable complete set of basis functions, denoted by $P_n$:

$$F(x_i) = \sum_{n=0}^{N} A_n P_n = 0 \ . \tag{2}$$

As our 3D codes for evolving the black hole spacetimes are written in cartesian coordinates, STF tensors are most convenient for our purpose. For relatively smooth surfaces, as one expects the AH to be, only the first few lower rank tensors (i.e., $N$ being finite and not too large, see below) are needed.

2. A surface is hence represented by the set of coefficients $A_n$. The expansion $\Theta$ at each point of the surface is a function of $A_n$. The condition for the trial surface to be the apparent horizon becomes

$$\int_\sigma \Theta^2(A_n) = 0 \tag{3}$$

As $\Theta^2(A_n)$ is semi-positive definite, there are efficient optimization algorithms to search the $A_i$-space for the surface closest to the AH among all test surfaces so parameterized.

## 2. Implementation and Results

We first applied our AH finder based on this method to 2D testbeds. We used as our background spacetime data obtained from a code developed by Bernstein *et al.*[3] This code evolves a black hole distorted by an axisymmetric distribution of gravitational waves (Brill wave). The black holes can be highly distorted by the incoming wave. We compare the results obtained with our new AH finder with the results from the AH finders constructed using the standard PDE method. For test surfaces with 16 coefficients (with the spherical harmonics $Y_{lm}$ as basis functions), we find that both the new method and the old methods coincide within the given accuracy of the PDE solvers. The new method takes less than 5 minimization iterations to converge to the correct surface in typical situations where we use the previous solution as a guess on the present time slice.

For the 3D testbeds, we use the symetric trace free (STF) tensors as basis functions. For the Schwarzschild spacetimes constructed in our 3D cartesian codes,[4] we find that out method can converge to the correct Schwarzschild horizon location within a few iterations and using only a 6 coefficient expansion. The horizon is correctly located to within a small fraction of a grid zone. Fig. 1 shows how the 3D AH finder converges. We are presently testing the AH finder on evolved Schwarzschild black holes and also on two black hole data sets in 3D cartesian coordinates. Results of this work will be reported elsewhere.

## 3. Conclusion

We have developed a promising method of finding the AH in a numerically constructed spacetime by a minimization procedure. This method should be an efficient and accurate tool for use in 3D numerical relativity.

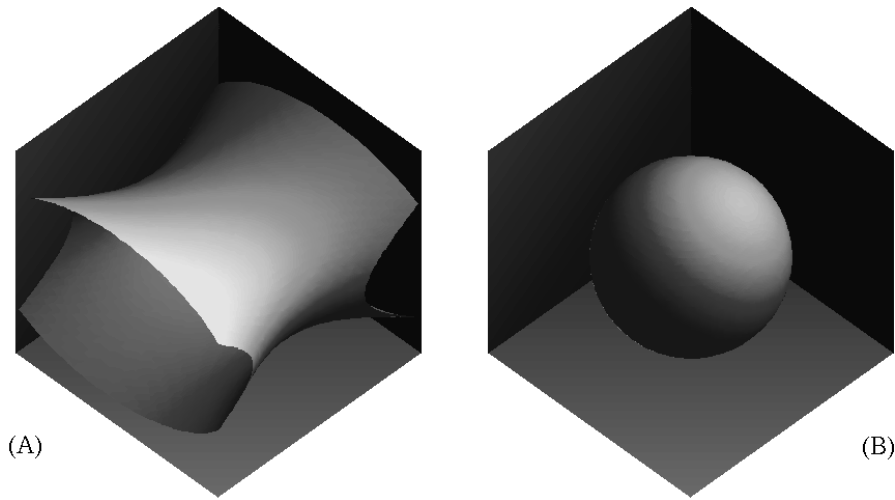

Figure 1: (A) Initial trial surface. (B) After 4 minimization iterations, we converge to the spherical Apparent Horizon for a Schwarzschild black hole in a 3D cartesian grid.

## Acknowledgements

This research is supported by the NCSA, the Pittsburgh Supercomputing Center, and NSF grants Nos. PHY91-16682, PHY94-04788, PHY94-07882 and PHY/ASC93-18152 (arpa supplemented). J.M. also acknowledges a Fellowship (P.F.P.I.) from Ministerio de Educación y Ciencia of Spain.

## References

1. G. Cook and J. W. York, Phys. Rev. D, **41**, 1077 (1990).
2. K.S. Thorne, Rev. Mod. Phys. **52**, 299 (1980).
3. D. Bernstein *et al.*, Phys. Rev. D, **50**, 5000 (1994).
4. P. Anninos, K. Camarda, J. Massó, E. Seidel, W.-M. Suen, M. Tobias, J. Towns. "3D Numerical Relativity at NCSA", in these proceedings.